# Gold nanorod induced enhanced efficiency in luminescent solar concentrator device

Puspendu Barik[1,a], Jaydeep Kumar Basu[1]

AFFILIATIONS

[1]Soft Nano Materials Physics Group, Department of Physics, Indian Institute of Science, Devasandra Layout, Bengaluru, 560012, India,
[a]Author to whom correspondence should be addressed: pbarik_mid1983@yahoo.co.in

ABSTRACT

We have observed significant changes in the edge emission of a luminescent solar concentrator device (LSC) consist of core shell $Cd_{1-x}Zn_xSe_{1-y}S_y$ quantum dots (QDs), and a monolayer of gold nanorods (GNRs) on the surface of LSC device. The observed changes show a nonlinear growth when another LSC of same thickness casted on the top of GNRs layer. The mechanism of plasmon-enhanced PL is mainly associated with the surface plasmon excitation which were found to be imprinted into the corresponding PL characteristics collected from the edge and back side of LSCs. The findings point to so far not recognized any application potentials of plasmonic LSC device.

To achieve a decent power conversion efficiency (PCE) from luminescent solar concentrator (LSC) devices, researchers are continuously trying to develop new strategies to incorporate different type QDs, altering the waveguide materials, varying the assembly procedure and so on.[1–4] Several articles have tried to exploit the broad plasmonic absorption band of metal nanoparticles (MNPs) to sensitize the luminescence of different luminophores by means of the surface plasmon resonance (SPR) near-field enhancement effect. Metal nanorods act as nanoantenna and can greatly enhance the emission efficiency of QD films.[5–7] The doping of certain amount of MNPs with the emitters shows the enhancement in PCE of Plasmonic LSC (PLSC) as reported by several authors.[8–11] PLSC provides higher PCE than common LSC devices and suitable to use in building-integrated photovoltaics (BIPV) systems due to its transparent/semi-transparent properties.[12] When a quantum emitter (e.g. QDs or dye molecules) is near to the enhanced local electric field intensity of gold nanorods (GNRs), plasmonic interaction takes place which can improve light absorption, the excitation rate, radiative and non-radiative decay rates of the optical emitter. In this current study, we report the enhancement of waveguiding efficiency of a conventional LSC device by casting a monolayer of GNRs on the top which serves as subwavelength scattering elements to couple and trap freely propagating plane waves from incident light into the absorbing waveguiding materials containing the quantum emitters by folding the light into the layer.[13] The results expand the potential applicability of such PLSC devices in unexplored territory, i.e., the realization of PLSC devices employing coupling and trapping effect of layer of GNRs, and the avenues of so far unexplored in vivo application possibilities for BIPV. The mechanism of plasmon-assisted light trapping in the device and the emission enhancements by SPR were found to be imprinted into the corresponding PL characteristics.

We have synthesized highly luminescent and stable $Cd_{1-x}Zn_xSe_{1-y}S_y$ QDs (from now on we would represent these by writing QDs only) with chemical composition gradients having fluorescent quantum yield (FQY) of ~82% on large scale via a single-step synthesis method by using the principle of reactivity difference between Cd and Zn precursors.[14] The QDs were characterized for their crystal structure and phase using X-ray diffractometer (XRD, Rigaku) with $\lambda$ = Cu-K$_\alpha$ (1.5418 Å) radiation source. The lattice planes are assigned with zinc-blend structure of CdSe and ZnS as shown in the Fig. S1 (Supporting Information). Transmission electron microscopy (TEM) images were obtained using a TecnaiTM G2 F30 S-Twin operating at an acceleration voltage of 300 keV, which possess predominately spherical shapes with a diameter of ~6.7 nm as shown in Fig. S2 (Supporting Information). GNRs were synthesized by using the methods as reported by El-Sayed group[15] and Liz-Marzán group[16] followed by a shortening process[17] of GNRs with low aspect ratio to achieve a longitudinal SPR (LSPR) band centered at 580 nm and a weak transverse SPR band as shown in Fig. 1. In a typical LSC device, QDs were dispersed at several concentrations in a mixture of lauryl methacrylate (LMA) with the cross-linking agent ethylene glycol dimethacrylate (EGDM) and the UV-initiator of diphenyl(2,4,6-trimethylbenzoyl)phosphine oxide by adapting a previously reported procedure[18] which is schematically in Fig. 2. For rigid device with a fixed thickness of 1 mm (with area of 2 cm×2 cm), we have selected the ratio of LMA:EGDM to 80:20 (w/w) and the amount of QDs used 0.1 wt. % (LSC-1), 0.2 wt. % (LSC-2) and 0.4 wt. % (LSC-3). A monolayer of GNRs was placed on the top of LSC device by following the method adapted by our group[19] using Langmuir–Schaefer (LS) technique (Kibron Micro trough, Finland) at room temperature (RT, 25 °C) and devices are designated here by PLSC-1, PLSC-2 and PLSC-3 respectively without compromising the transparency (~70 - 80%).

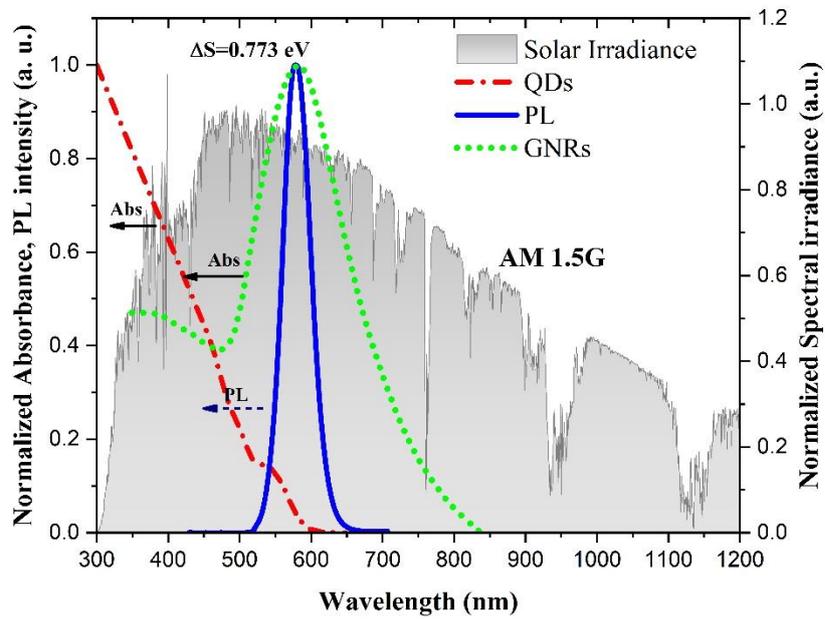

*Figure 1 Absorption (Abs) and photoluminescence (PL) spectra of QDs and GNPs used in this article with the AM 1.5G solar spectrum obtained from the national renewable energy laboratory (NREL) (grey shades). The direct optical band gap (Eg) for QDs is extracted from Tauc plot and ΔS=773 meV, represent the Stoke shift for QDs arises due to graded shell of QDs*

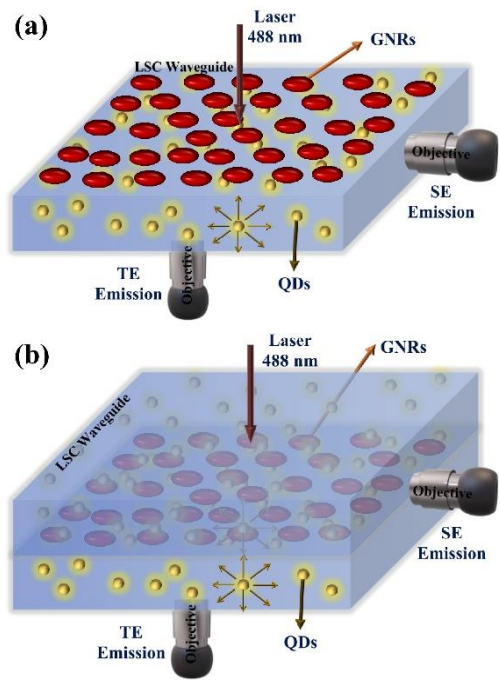

*Figure 2 (a) shows the device structure and measurement configuration used in the article where one monolayer of GNRs were casted by LS technique on the top of LSC device containing QDs with different concentration, (b) shows sandwiched like device structure where one LSC device of same thickness as below casted on the top of GNRs layer.*

UV-visible (UV-Vis) spectra were recorded at RT using UV/Vis/NIR spectrophotometer (LAMBDA™ 750, Perkin Elmer) and PL spectra for solution/film were recorded using luminescence spectrometer (LS55, Perkin Elmer) which shows no significant changes in the optical spectra of bare QDs film and the film of QDs inside waveguide. The quantum efficiency (QE) of QDs in LMA-EGDM polymer composite [P(LMA-co-EGDM)] is somewhat higher than in the solution/film which is suggesting a better stabilization of QDs in the polymer matrix confirmed by comparing the absorption and PL in toluene and in P(LMA-co-EGDM). In general, for P(LMA-co-EGDM) waveguides with refractive index n ≈ 1.49, the maximum light trapping efficiency $\eta_{TR} = \sqrt{1 - 1/n^2} = 74.13\%$, resulting in ~25% optical loss at any reabsorption/re-emission event and Fresnel losses will not exceed ~4%. The plasmonic interaction was studied through measured photoluminescence (PL) of QDs/GNRs structure quantifying the side emission (SE) and transmitted emission (TE) as depicted in Fig. 2. The emission spectra from LSC devices were recorded with the Confocal Raman imaging microscope (WITec alpha300 RS, excited with 488 nm laser and power density <100 W/cm$^2$) for TE and SE spectra which is schematically presented in Fig. S3 (Supporting Information). We have calculated the ratio of the SE and TE after normalization of collection efficiency of objectives used in the experiments. All the results, related to SE/TE ratio as shown in the Fig. 4, are averaged from several experiments changing the measurement side of LSC device as shown in the Fig. 2 (keeping the excitation at the center of LSC plane).

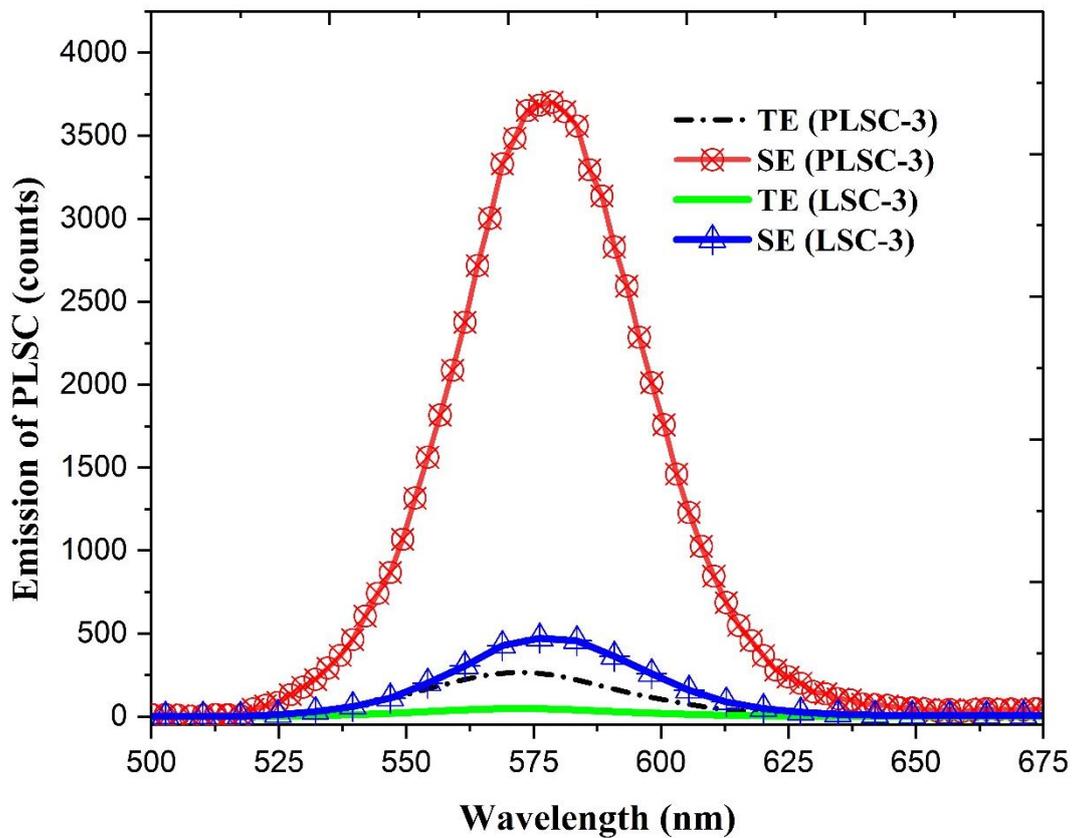

*Figure 3 The figure shows representative SE and TE emission for LSC-3 and PLSC-3 devices.*

PLSC-3 shows a maximum SE/TE ratio of ~15 with GNRs compared to ~10 for LSC-3 as shown in inset of Fig. 4. To enhance further, we have tested sandwiched like structure of PLSC as shown in the Fig. 2(b), where a layer of LSC of same thickness have been casted on the top of the monolayer of GNRs. The device shows more than 100% enhancement of the ratio of SE/TE (Fig. 4). For reference, we have measured LSC device with thickness of 2 mm without any GNR layer which shows ~30-36% increase in SE/TR ratio than LSC device with 1 mm thickness as shown in Fig. 4. The reasons behind the enhancement of edge emission are categorized into three different effect - light scattering due to plasmons, light concentration because of strong local field enhancement around the GNRs to increase absorption, and light trapping using surface plasmon polaritons (SPP). All these effects ensure that light must travel to edge more without significant loss in PLSC device than LSC. First, light scattered from the layer of GNRs at an angle beyond the critical angle for reflection (42° for the P(LMA-co-EGDM)/air interface) will remain trapped in the device. In addition, light traveled towards the back surface will absorb and scatter by the QDs inside LSC (1 mm thick) and be partly reradiated into the LSC by the same scattering mechanism. As a result, the incident light will pass several times through the LSC waveguide due to repeated absorption, reabsorption and scattering events increasing the effective path length. Second, GNRs are acting as an effective

'nanoantenna' for the incident light which store energy in the form of a localized surface plasmon mode. In our case, as shown in Fig. 1, LSPR bands overlaps with the emission spectrum of QDs which leads to effective coupling of LSPR to far-field radiation ensuing an increased emission rate and quantum yield. Third, incident light is transformed into SPPs and travel along the interface between the layer of GNRs and LSC waveguide which are confined near the interface at dimensions much smaller than the wavelength of incident light. SPPs, effectively turned the incident light by 90°, thus trap and guide light in the LSC. Our results corroborate the explanations stated above which are depicted in Fig. 4 with more SE/TE ratio for sandwiched type structure of PLSC device.

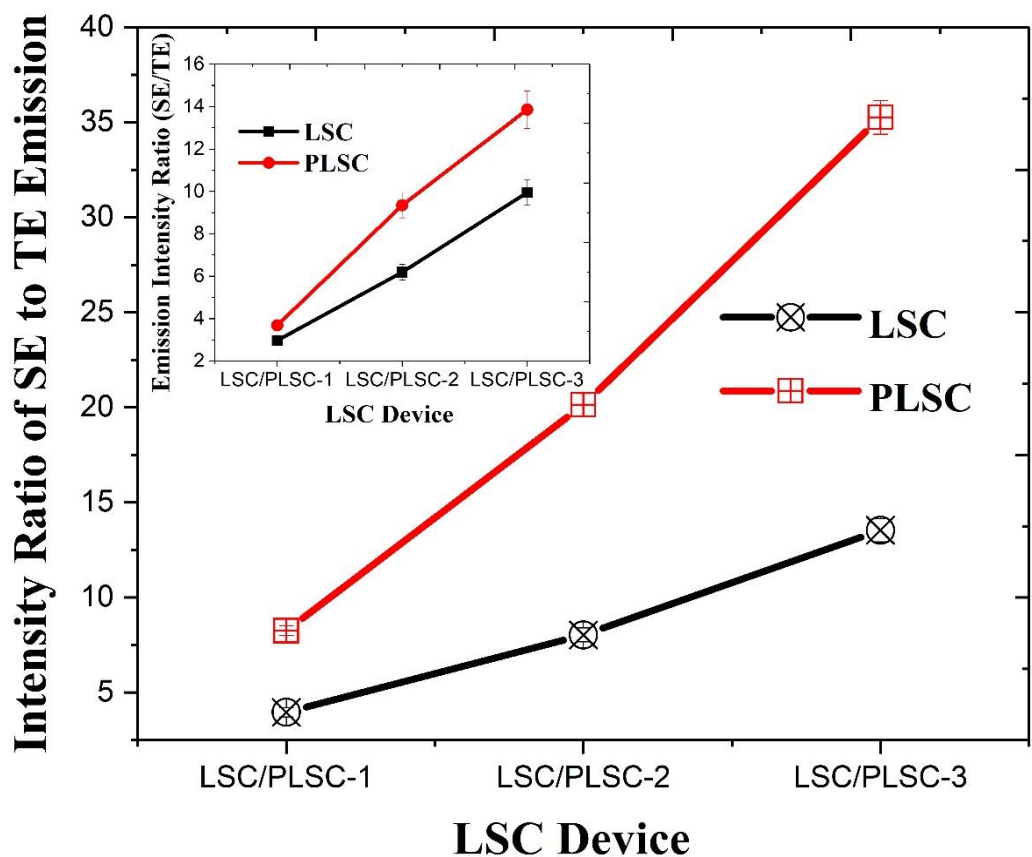

*Figure 4 The figure shows the emission intensity ratio of SE/TE for three LSC devices with and without GNRs for a device with a monolayer of GNRs with sandwiched between two LSCs. The inset of the figure shows the intensity ratio of SE/TE emission from LSC and PLSC devices with a GNRs layer on the top.*

Currently under exploration, but important to note with respect to technological exploitations of the here-in reported phenomena, is the finding that the insertion of GNR layer with conventional LSC devices provoke emission from side about 50% or more. Hence, the realization of PLSC with layer structure of GNRs seems possible and merits further investigation. Summarizing, the incorporation of GNRs layer would enhance the waveguiding efficiency to certain extend so that it would help to realize commercialized PLSC device in future. The observations open so far unexplored avenues for plasmonic enhancement of emission in PLSC device, getting more emission at side, and alternative approach to improve the PCE of the device. We anticipate the findings reported will stimulate more theoretical treatments for the improvement of efficiency of PLSC device.

## SUPPORTING INFORMATION
See the supporting information for the powder XRD, TEM images of QDs and the schematic which represents the measurement configuration in WiTec confocal microscope for RE/TE of LSC devices.

## ACKNOWLEDGEMENT
*University Grant Commission (UGC), New Delhi, India provided support for this research under Dr. D S Kothari Postdoctoral fellowship scheme. We acknowledge the Department of Science and Technology (DST), India for the financial support and Advanced Facility for Microscopy and Microanalysis, Indian Institute of Science, Bangalore. Authors Acknowledge Riya Dutta for her help in taking TEM image of CdSe/ZnS QDs.*

# Supporting Information

# Gold nanorod induced enhanced efficiency in luminescent solar concentrator device

Puspendu Barik[1], Jaydeep Kumar Basu[1]

[1]Soft Nano Materials Physics Group, Department of Physics, Indian Institute of Science, Devasandra Layout, Bengaluru, 560012, India

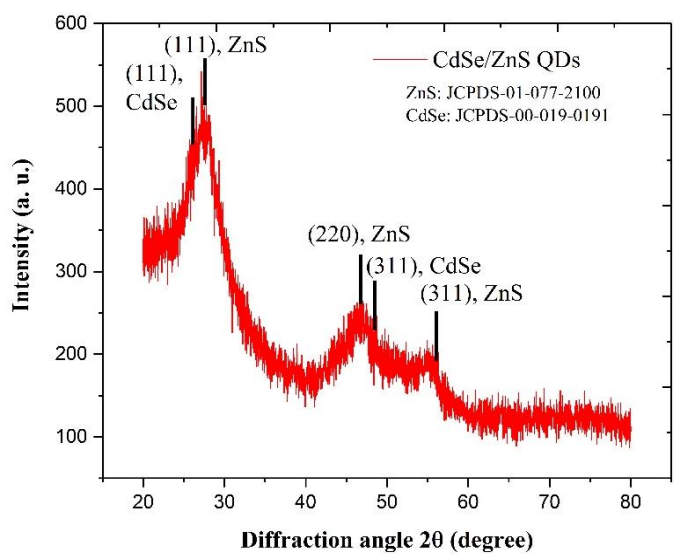

*Figure S1*. XRD patterns of QDs capped with oleic acid (OA) and shows the diffraction peak positions for CdSe (JCPDS- 00-019-0191) and ZnS (JCPDS- 01-077-2100). The positions of the diffraction peaks standing for different planes referred to the zinc blend structure of CdSe and ZnS

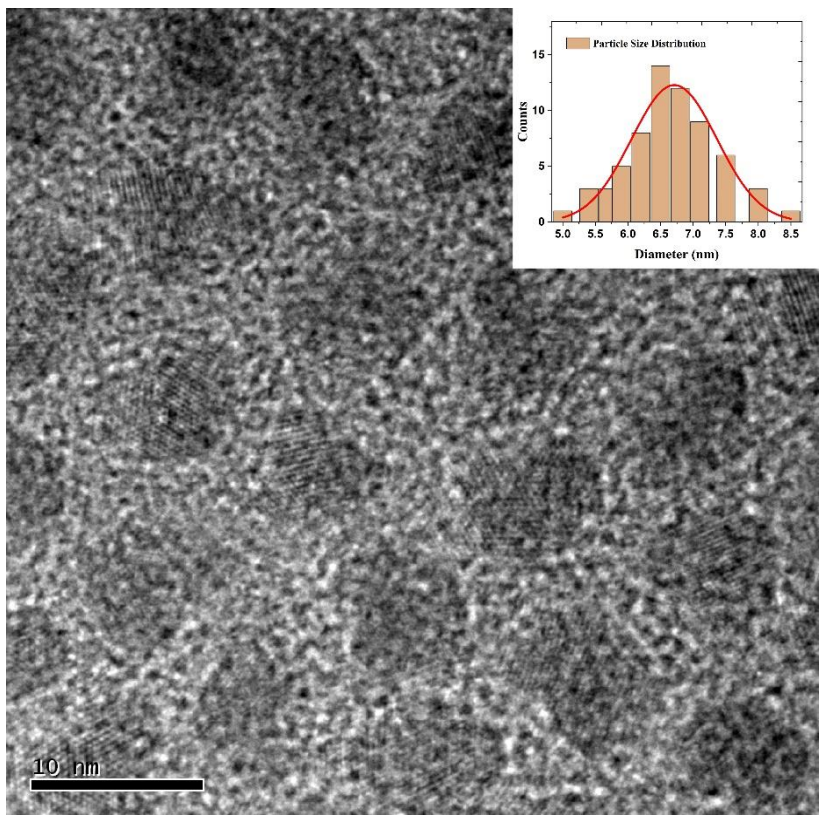

*Figure S2. Representative TEM image of QDs and inset shows the histogram of the corresponding diameter of the QDs as analyzed by TEM using the Gatan Microscopy Suite (GMS) software. The solid line represents a Gaussian distribution fit to the data showing the average size of QDs around ~6.7 nm*

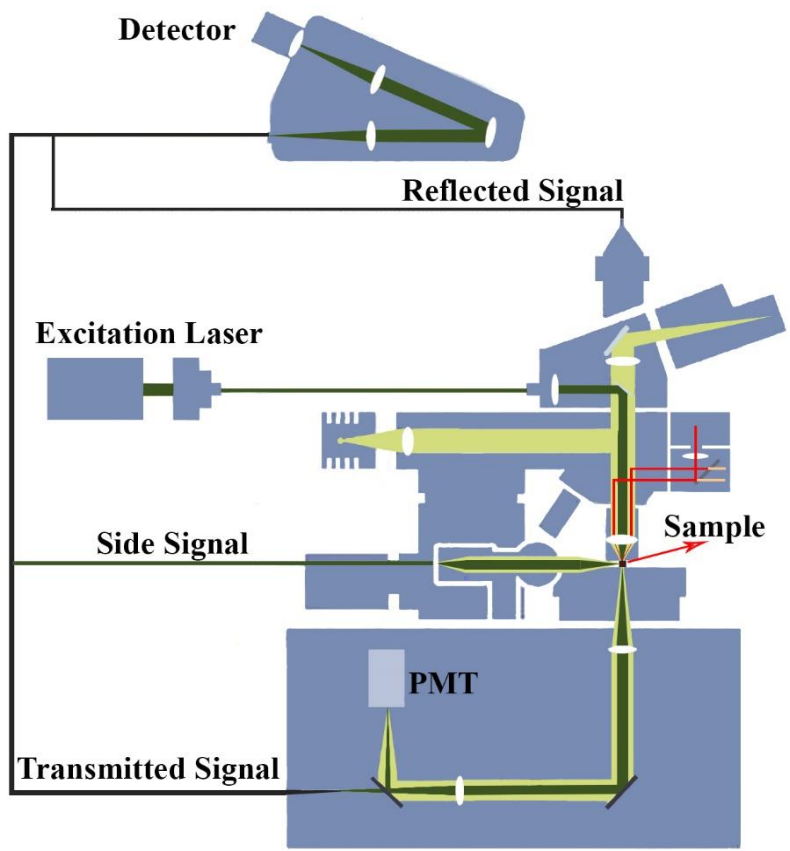

***Figure S3***. *the schematic represents the measurement configuration of confocal microscope (WiTec alpha 300 RS) and indicating the signals depicted in the text.*